\documentclass[aps,prl,showpacs,preprint,floatfix]{revtex4}
\usepackage{latexsym}
\usepackage{amsfonts}
\usepackage{amssymb}
\usepackage{amsmath}
\usepackage{graphicx}

\begin{document}

\title{High domain wall velocities due to spin currents perpendicular to the plane}

\author{A. V. Khvalkovskiy} \altaffiliation{Corresponding author: khvalkov@fpl.gpi.ru} \affiliation{General Physics Institute of RAS, 119991 Moscow, Russia and Unit\'e Mixte de Physique CNRS/Thales and Universit\'e Paris Sud 11, 91767 Palaiseau France} 
\author{K. A. Zvezdin, Ya. V. Gorbunov}
\affiliation{General Physics Institute of RAS, 119991 Moscow, Russia}
\author{V. Cros, J. Grollier}\affiliation{Unit\'e Mixte de Physique CNRS/Thales and Universit\'e Paris-Sud 11, 91767 Palaiseau France}
\author{A. K. Zvezdin}
\affiliation{General Physics Institute of RAS, 119991 Moscow, Russia}

\date{\today}

\begin{abstract}
We consider long and narrow spin valves composed of a first magnetic layer with a single domain wall (DW), a normal metal spacer and a second magnetic layer that is a planar or a perpendicular polarizer. For these structures, we study numerically DW dynamics taking into account the spin torques due to the perpendicular spin currents. We obtain high DW velocities: 50 m/s for planar polarizer and 640 m/s for perpendicular polarizer for $J$ = $5 \times 10^6$ A/cm$^2$. These values are much larger than those predicted and observed for DW motion due to the in-plane spin currents. The ratio of the magnitudes of the torques, which generate the DW motion in the respective cases, is responsible for these large differences. 
\end{abstract}

\pacs{75.60.Ch,72.25.Ba,85.75.-d}\maketitle

The control of the dynamics of a geometrically confined magnetic domain wall (DW) has attracted much attention in the last decade because of the fundamental interest to identify some specific properties of such a nanoscale spin structure and also for the promising applications such as DW-based magnetic memory and logic devices. Recently, field induced DW motion in magnetic nanostripes has been extensively studied \cite{Atkinson, Beach, ThiavFDDWM}. However this solution is hardly transferable for dense arrays of submicronic devices. An alternative to overcome this challenging task of addressing properly a magnetic bit is to use current induced domain wall motion, that has been the subject of many experimental \cite{Marrows, Grollier, Yamaguchi, Klaui, Parkin, ravelosona} and theoretical researches \cite{Bazaliy, Tatara, LZprb, ZLprl, ThiavEpl}. The main intrinsic mechanism to bring the DW into motion by the current is the spin transfer torque, resulting from a transfer of a spin momentum from the conduction electrons to the local spin system inside the DW \cite{berger}. 

In most cases, the nanostripe consists of a single magnetic film (e.g. NiFe) and thus the spin polarized current is flowing in the plane of a magnetic layer that contains the DW: the so-called current-in-plane (CIP) configuration. Our purpose is to study numerically the DW motion for the case of spin currents flowing perpendicular to the plane. Our results show that for this current perpendicular to the plane (CPP) configuration, a DW can be set into a steady-state motion with very high velocities, up to two orders in magnitude larger than those for a CIP stripe with similar applied current densities. 

Multidomain states excited by the spin currents in CPP spin valve nanopillars have been already discussed in literature \cite{acreman,mangin,rebei}. For example, strongly non-uniform magnetic states were observed in the such nanopillars switched by the current\cite{acreman}. Static two-domain states were observed in CPP nanostructures with strong perpendicular magnetic anisotropy \cite{mangin}. Current-induced dynamics of a trapped DW in a square nanopillar was numerically studied in Ref. \cite{rebei}. However, for such small pillars, the motion of a DW is impeded due to the strong interaction of the DW with the edges or with other DWs. In contrast to this previous work, we consider a long and narrow CPP spin-valve structure with a single DW. This original structure allows us to investigate the DW motion subjected only to the CPP-spin transfer torque. 

\begin{figure}[h]
   \centering
    \includegraphics[keepaspectratio=1,width=6.5 cm]{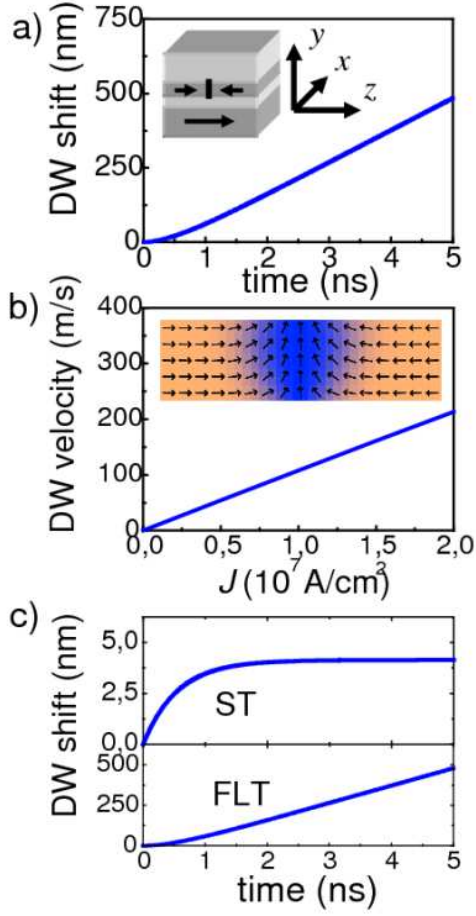}
     \caption{ (color online) (a) DW displacement for $J$ = $1 \times 10^{7}$ A/cm$^{2}$ for a CPP spin valve with the planar polarizer. Inset: sketch of the device. (b) Velocity of the steady-state DW motion as a function of the current density $J$. Inset: shape of the DW during its steady-state motion for $J$ = $1 \times 10^{7}$ A/cm$^{2}$; the color represents the x-component of magnetization ranging from 0 to + $M_{s}$. (c) DW displacement taking into account only the Slonczewski torque ST (top panel) or only the field-like torque FLT (bottom panel).}
    \label{fig1}
\end{figure}

\begin{figure}[h]
   \centering
    \includegraphics[keepaspectratio=1,width=6.5 cm]{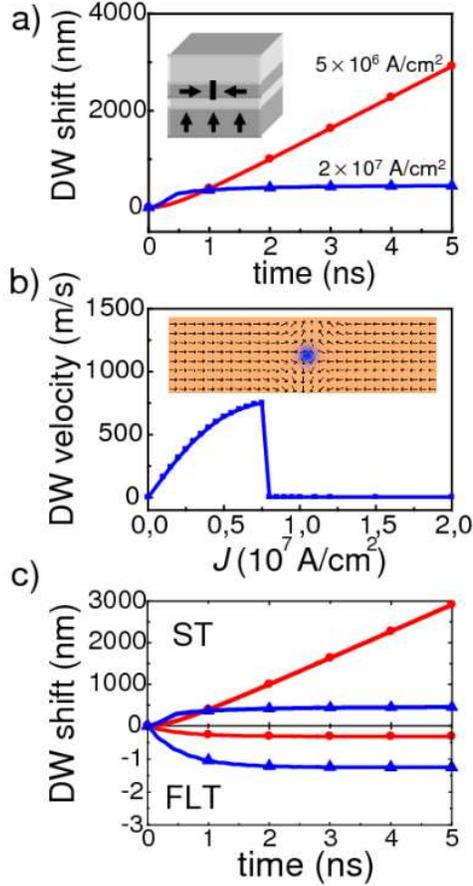}
     \caption{ (color online) (a) Current-induced DW displacement for $J$ = $0.5 \times 10^{7}$ A/cm$^{2}$ (marked by circles) and for $J$ = $2 \times 10^{7}$ A/cm$^{2}$ (marked by triangles) for a CPP spin valve with the perpendicular reference layer. Inset: sketch of the device. (b) Velocity of the steady-state DW motion as a function of the current density J. Inset: snapshot of the transformed and eventually stopped DW (calculated for $J$ = $2 \times 10^{7}$ A/cm$^{2}$); the color represents the out-of-plane component of magnetization ranging from 0 to + Ms. (c) DW displacement taking into account only the Slonczewski torque ST (top panel) or only the field-like torque FLT (bottom panel), for $J$ = $0.5 \times 10^{7}$ A/cm$^{2}$ and for $J$ = $2 \times 10^{7}$ A/cm$^{2}$.}
    \label{fig2}
\end{figure}

In the simulations, the magnetic stack is the following: a reference magnetic layer, a non magnetic spacer and a free magnetic layer. The free layer has an in-plane magnetization and contains a single DW. The reference layer has a single domain fixed magnetization. We have considered two magnetic configurations for the reference layer: a reference layer magnetized in the plane of the film, referred further to as a planar polarizer, and a reference layer magnetized perpendicularly to the plane, a perpendicular polarizer (see the insets of Fig.\ref{fig1}(a) and \ref{fig2}(a)). We assume that the current flows perpendicularly to the layers and has a uniform distribution. The magnetization dynamics is described by the modifed Landau-Lifshitz-Gilbert (LLG) equation taking into account the spin transfer effect \cite{Slonc}:
\begin{eqnarray}\label{LLG}
\frac{d\textbf{M}}{dt} & = & -\gamma \textbf{M} \times \textbf{H}_{eff} + \textbf{T}_{STT} + \frac{\alpha }{M_{s}}\left(  \textbf{M} \times \frac{d\textbf{M}}{dt} \right) 
\end{eqnarray} 
where \textbf{M} is the magnetization vector, $\gamma$ is the gyromagnetic ratio, $M_{s}$ is the saturation magnetization,  $\alpha$ is the Gilbert damping. The effective field $\textbf{H}_{eff}$ is the sum of the the magnetostatic field, the exchange field and the anisotropy field. The spin transfer torque $\textbf{T}_{STT}$ is divided in two components often referred to as a Slonczewski torque $\textbf{T}_{ST}$ (ST) and a field-like torque $\textbf{T}_{FLT}$ (FLT) \cite{ZLF, Stiles, Bauer}: 
\begin{eqnarray}\label{ST}
\textbf{T}_{ST} & = & -\gamma \frac{a_{J} }{M_{s}} \textbf{M} \times \left[\textbf{M} \times \textbf{m}_{ref} \right]
\end{eqnarray} 
\begin{eqnarray}\label{FLT}
\textbf{T}_{FLT} & = & -\gamma b_{J} \left[\textbf{M} \times \textbf{m}_{ref} \right]
\end{eqnarray} 
where $\textbf{m}_{ref}$ is a unit vector along the magnetization direction of the reference layer. The ST amplitude is given by $a_{J}$ = $\hbar J P / 2 d e M_{s}$, where $d$ is the thickness of the free layer, $J$ is the current density, $e>0$ is the charge of electron and $P$ is the spin polarization of the current. The amplitude of the FLT is $b_{J}$ = $\xi_{CPP} a_{J}$, where $\xi_{CPP}$ is typically about 0.1 \cite{Stiles, Zimmler}. In the following, we will also consider the two effective fields related to the two torques $\textbf{T}_{ST}$ and $\textbf{T}_{FLT}$: $\textbf{H}_{ST}$ = $\frac{a_{J} }{M_{s}} \left[\textbf{M} \times \textbf{m}_{ref} \right]$ and $\textbf{H}_{FLT}$ = $ b_{J} \textbf{m}_{ref}$.

For the calculations, we consider that the reference layer is a fixed spin polarizer and therefore we study the dynamics only in the free layer. The initial magnetic configuration in the free layer is a transverse head-to-head DW wall located in the middle of the stripe. The current is switched on at time t = 0 with a zero rise time. The magnetic parameters used in the simulations are for Co: $M_{s} =$ 1400 emu/cm$^{3}$, the exchange constant A = 1.3 $\times$ 10$^{-6}$ erg/cm, $\alpha = 0.007$, and we neglect the bulk anisotropy. We assume that the free layer has perfect edges and is 50 $\times$ 3 $\times$ 8000 nm$^{3}$ in size. For both studied structures that is the planar or perpendicular polarizer, we have used a spin polarization P = 0.32 (corresponding to $a_{J}$ = 25 Oe at $J$ = 10$^{7}$ A/cm$^{2}$) and $\xi_{CPP}$ = 0.1. To compare the efficiency of each torque in the current-induced domain wall motion we perform some additional simulations, in which only one torque, the ST or the FLT, is taken into account. 

The simulations consist in a numerical integration of the LLG equation (1) on a two-dimensional mesh using our home-made micromagnetic code using the forth order Runge-Kutta method with an adaptive time-step control for time integration. The mesh size is 2.5 nm. The DW position and velocity are extracted from the average magnetization of the whole element. The Oersted field and the thermal fluctuations are not taken into account here. We emphasize that the injected current densities are always below the threshold values corresponding to the reversal or the excitation of the magnetization inside the domains adjacent to the DW.

In Fig.\ref{fig1}, we present the results for the planar polarizer. In Fig.\ref{fig1}(a), we show the calculated DW displacement for $J$ = $1 \times 10^{7}$ A/cm$^{2}$. The DW moves immediately after applying the current and then reaches a regime of steady motion after about 2 ns. In this latter regime, the DW velocity is 105 m/s.  We find that the DW velocity increases linearly with the current density J as shown in Fig.\ref{fig1}(b) (the threshold current density for domain excitation due to the spin transfer is $J$ = $2.4 \times 10^{7}$ A/cm$^{2}$). Note that the shape of the DW, displayed in the inset to Fig.\ref{fig1}(b), remains virtually unchanged during the steady motion. Looking at the separate treatment of the torques $\textbf{T}_{ST}$ and $\textbf{T}_{FLT}$ in Fig.\ref{fig1}(c), it is obvious that it is the FLT that is responsible for the steady DW motion. The effect of the ST is only to shift the DW at a small finite distance of about 4 nm. We emphasize that this displacement is almost negligible compared to the hundreds nanometers on which the DW travels during the first few nanoseconds due to the FLT. 

A simple explanation of these results can be given by analyzing the symmetry of the different torques \cite{ThiavFDDWM, LZprb, ThiavEpl}. The magnetization dynamics according to Eq. (1) is a gyration around the effective field and a motion towards the effective field due to the damping. For the ST, $\textbf{H}_{ST}$ is directed perpendicular to the plane. The gyration around $\textbf{H}_{ST}$ leads to the DW displacement. Due to the damping, the magnetization tilts towards $\textbf{H}_{ST}$. As a result magnetic charges appear at the opposite sides of the free layer. They generate a magnetostatic field which balances $\textbf{H}_{ST}$, thus the DW stops. $\textbf{H}_{FLT}$ is a uniform field directed along $\textbf{m}_{ref}$, thus the DW moves due to the FLT like as it is expected for a uniform magnetic field. Using the conventional framework (e.g. similarly to Ref. \cite{LZprb}), it is possible to write an analytical equation for the DW motion in a one-dimensional approximation: 
\begin{eqnarray}\label{RWA}
\frac{u}{\Delta} -\alpha \dot{\Phi}+\gamma a_{J} & = & \frac{\gamma}{M_{s}} K_{x} \sin (2 \Phi) \nonumber \\ 
- \alpha \frac{u}{\Delta} + \dot{\Phi} + \gamma b_{J} & = & 0
\end{eqnarray}
Here the collective coordinates  $(u,\Phi)$ define the DW with a traveling wave ansatz $\theta \left(z,t\right)$ = $2 \tan^{-1}\left(\frac{z-ut}{\Delta}\right)$, $\phi \left(z,t\right)$ = $\Phi \left(t\right)$, where $(\theta, \phi)$ are the polar angles defining the magnetization direction; $\Delta$ is the parameter of the domain wall width, $K_{x}$  is the shape anisotropy constant for the $x$-axis. The steady-state solution of Eq. (4) is $\dot{\Phi}=0$, with the DW velocity given by $u$ = $\frac {\gamma b_{J} \Delta}{\alpha}$. This analytical result is in a very good agreement with our numerical findings. It shows that the final DW velocity does not depend on $a_{J}$ and gives $\Delta$ = 17 nm by fitting the data in Fig.\ref{fig1}(b) vs. $\Delta$  = 19 nm yielded from the fit of the magnetization profiles to the traveling wave ansatz.

The results for the case of the perpendicular polarizer are presented in Fig.\ref{fig2}. We clearly identify two behaviors for the DW motion for the current densities below or above the critical value $J_{c}$ = $0.8 \times 10^{7}$ A/cm$^{2}$. For $J < J_{c}$, as we show for $J$ = $0.5 \times 10^{7}$ A/cm$^{2}$ in Fig \ref{fig2}(a), the behavior is similar to the previous planar case, i.e. after a short transient phase, the DW moves steadily. The steady-state DW velocity is here 640 m/s  for $J$ = $0.5 \times 10^{7}$ A/cm$^{2}$, and it increases with the current density as shown in Fig.\ref{fig2}(b). An interesting result is that, at a given current density, the DW velocity is much higher than in the case of a planar polarizer. We find that the ratio of the DW velocities ranges from 15 for $J << J_{c}$ to 9.4 for $J = J_{c}$. For the large-current regime ($J \geq J_{c}$), the DW undergoes a structure transformation from a transverse DW to an antivortex DW, which strongly influences the DW motion (see the simulation results for $J$ = $2 \times 10^{7}$ A/cm$^{2}$ in Fig \ref{fig2}(a)). We observe the formation of an antivortex at one of the edges of the stripe, which is then shifted towards the middle (see the resulting magnetization distribution the inset to Fig.\ref{fig2}(b)). Whereas the antivortex approaches the center, the DW velocity diminishes until it eventually vanishes. Contrary to the planar case, the DW steady motion is generated by the ST, as we conclude by analyzing the simulation results for a separate treatment of the two torques (see Fig.\ref{fig2}(c)). The action of the FLT is to shift the DW on a very small distance (0.3 nm for $J$ = $0.5 \times 10^{7}$ A/cm$^{2}$) in the opposite direction.   

In this case of the perpendicular polarizer, $\textbf{H}_{ST}$ points in the film plane along the magnetization direction of one of the two domains adjacent to the DW. This induces a steady DW motion. $\textbf{H}_{FLT}$ is directed perpendicularly to the plane, thus the FLT results only in a finite displacement of the DW. The magnetization gyration around $\textbf{H}_{ST}$ is a rotation towards perpendicular to the plane direction. This rotation is balanced by the demagnetization field in the low-current regime ($J < J_{c}$), however for the larger currents ($J \geq J_{c}$), it results in the formation of the antivortex. Once the antivortex reaches the middle line of the stripe, the total torque acting on the DW vanishes and the DW stops \cite{analog}. The large difference for the steady-state DW velocities (at low currents) between the perpendicular polarizer  and the planar polarizer is linked to the multiplication factor $\xi_{CPP}$ between the torques. 

Hereafter we compare our simulation results to those obtained for the spin transfer induced DW motion in the CIP systems, such as a nanostripe made of a single magnetic material \cite{Bazaliy, Tatara, LZprb, ZLprl, ThiavEpl}. In these systems, the effect of the spin transfer is usually introduced as an additional torque $\textbf{T}_{STT}$ in the LLG equation similarly to Eq. (1) for the CPP case. In the most recent model, the spin torque has been divided in two components. A first one, called the adiabatic torque $\textbf{T}_{a}$, takes into account the spin transfer from spins of the conduction electrons that perfectly track the local magnetization inside the DW, and the second nonadiabatic torque $\textbf{T}_{n.a.}$ is a correction due to the mistracking of the electrons' spin from the non-uniform local magnetization. They are given by the expressions:
\begin{eqnarray}\label{Ta}
\textbf{T}_{a}  =  - \frac{ \gamma  c_{J}}{M^{2}_{s}} \textbf{M} \times \left[\textbf{M} \times ( \textbf{j}_{e} \nabla ) \textbf{M} \right]
\end{eqnarray} 
\begin{eqnarray}\label{Tna}
\textbf{T}_{n.a.} = - \frac{\gamma \xi_{CIP} c_{J} }{M_{s}} \left[\textbf{M} \times (\textbf{j}_{e} \nabla ) \textbf{M} \right],
\end{eqnarray} 
where $\textbf{j}_{e}$ is a unit vector along the current lines and, using our notations, $c_{J}=a_{J} d$, $d$ is the thickness of the magnetic layer \cite{Bazaliy, Tatara, LZprb}. The non-adiabaticity parameter $\xi_{CIP}$ is typically  0.001-0.05 \cite{ZLprl, ThiavEpl}. For metallic magnetic nanostripes, the adiabatic torque $\textbf{T}_{a} $ is predominant in terms of amplitude but it induces only a finite shift of the DW and a steady-state DW motion is due to the non-adiabatic spin torque. In this case, the DW velocity is given by $u$ = $\xi_{CIP} \gamma c_{J} / \alpha$  \cite{ZLprl}. From this expression it follows that the ratio of the predicted DW velocities for the CPP structure with the planar polarizer and the CIP magnetic stripe, at a given current density, is $\xi_{CPP} \Delta / \xi_{CIP} d$, that is typically a factor of 10 \cite{note}. Moreover, the same ratio considering now a CPP structure with the perpendicular polarizer and the CIP stripe is $ \Delta / \xi_{CIP} d$, that is about 100. Indeed, the estimation of the DW velocity at $J$ = $1 \times 10^{6}$ A/cm$^{2}$ is $u$ = 1 m/s for a CIP sripe with $\xi_{CIP} =$ 0.05 and the calculations give us $u$ = 11 m/s for the CPP structure in case of a planar polarizer and $u$ = 160 m/s in case of perpendicular polarizer. This clearly evidences the usefulness of vertical spin current to move a DW with a high efficiency in a nanostripe. Similar results are obtained for other materials, such as NiFe and CoNiPt (with large perpendicular anisotropy). Details of this study will be given elsewhere \cite{akhv}. We also note that the effect of the Slonczewski torque on a trapped DW was also discussed in Ref. \cite{rebei} for a CPP nanopillar, but a large-scale DW motion was not studied.

In the present calculations, we have assumed a uniform current flow over the whole cross section of the CPP spin valve structure. However, due to the local nature of the spin transfer effect, the most important contribution of the vertical spin currents comes from the spins that are going through the DW; indeed in the bulk of the domains the spin transfer torque vanishes. In fact, such situation with a localized vertical spin current in the vicinity of the DW could happen in a spin valve stripe with the current injected in plane \cite{Grollier}. In these CIP spin valve devices, the critical current densities for DW motion at zero field has been found to be a few $10^6$ A/cm$^2$, compared to about a few $10^7$ A/cm$^2$ in standard NiFe CIP stripes \cite{Yamaguchi}. The objective of this paper is not to give a quantitative explanation of these significant differences, but it is obvious that the spin currents flowing perpendicular to the plane beneath the DW might play an important role to explain the reduction of the critical current density for DW depinning. 

Our main findings may be summarized in comparing the predicted DW velocities for three different structures: a CIP magnetic nanostripe, a long CPP spin valve with a planar polarizer or a perpendicular polarizer. Different torques are responsible for the steady-state DW motion in these structures: the non-adiabatic torque, the field-like torque and the Slonczewski torque, respectively. The ratio of the magnitudes of the torques is the cause of the large difference in the DW velocities for these structures, approximately 1:10:100. The very effective DW motion for CPP structures we predict, could potentially be an original solution to achieve the necessary breakthrough in reducing the power consumption for the writing processes in spintronic devices.

We acknowledge A. Fert, A. Anane, S. Laribi, C. Chappert, S.S.P. Parkin and M. Viret for encouraging and helpful discussions. The work is supported by RFBR (grant 07-02-91589), RTRA "Triangle de la Physique", and ANR-07-NANO-034-04 "Dynawall".


\begin{thebibliography}{1}
\bibitem{analog}
This effect is similar to what is observed for field- and CIP current driven DW motion in magnetic nanostrips, see e.g. Ref. 3,13 . However in these latter cases, in contrast to our configuration, the torque acting on the DW does not vanish if the antivortex is in a middle of the strip, and therefore the DW does not stop its motion.
\bibitem{note}
Here and below it is assumed that the magnetic stripe of a CIP structure and the free layer stripe for a CPP spin valve are identical
\end{thebibliography}
\end{document}